\documentclass[10pt,a4paper]{article}
\usepackage[utf8]{inputenc}
\usepackage{amsmath}
\usepackage{amsfonts}
\usepackage{amssymb}
\usepackage{graphicx}
\author{Christoph Salge and Daniel Polani}
\title{Changing the Environment based on Intrinsic Motivation}

\begin{document}
\maketitle

One of the remarkable feats of intelligent life is that it restructures the world it lives in for its own benefit. Beavers build dams for shelter and to produce better hunting grounds, bees build hives for shelter and storage, humans have transformed the world in a multitude of ways. Intelligence is not only the ability to produce the right reaction to a randomly changing environment, but is also about actively influencing the change in the environment, leaving artefacts and structures that provide benefits in the future.

In this abstract, I want to explore if the framework of intrinsic motivation can help us understand and possibly reproduce this phenomenon. In particular, I will show some simple, exploratory results on how empowerment, as one example of intrinsic motivation, can produce structures in the environment of an agent. 

The basic idea behind intrinsic motivation is that an agent's behaviour, or decision making, is not guided by some form of externally specified reward, but rather by the optimization of some agent-internal measurement, which then gives rise to complex and rich interactions with the world. A quantitative formulation of an intrinsic motivation should ideally be computable from an agent's perspective, be applicable to different sensory-motoric configurations, and should reflect different agent embodiments. One classic example of intrinsic motivation is Schmidhuber's \cite{schmidhuber1991curious} artificial curiosity, where an agent acts in a way so that it learns the most about the environment. Other examples include Homeokinesis \cite{der1999homeokinesis}, or the predictive information framework \cite{Ay2008}. 

Here we want to focus on empowerment \cite{klyubin05:_all_else_being_equal_be_empow}, formally defined as the channel capacity between an agent's actuators, and its own sensors at a later time. This measures the potential causal flow in an agent's action-perception loop, and can be thought of as an abstract measure of how much control an agent has over the world it can perceive. The more meaningful options an agent has to influence the world, the higher is its empowerment. 

In a first, exploratory example, we applied discrete, 15-step empowerment to a three dimensional grid world. The model is somewhat inspired by the computer game ``Minecraft'' \cite{minecraft}. Each location is a cube, identified by its three integer coordinates. Each square can either be \textit{empty}, filled with \textit{earth}, or filled with an \textit{agent}. Each turn the agent can take one action. The agent can, for example, decide to move in the four cardinal directions, called \textit{north},\textit{ east}, \textit{south} and \textit{west.} If the cube is empty, the agent will enter it. If the cube is filled, then there are two options. If the cube above the target cube is empty, the agent will move in its desired direction and up by one unit. If the cube above is also filled, then the agent's move is blocked, and the agent will not move. If the agent afterwards is above an empty square, it will fall down one square. Earth blocks are not affected by gravity, so it is possible to have ``levitating'' structures. 

The agent can also decide to interact with the 6 adjacent blocks (\textit{up, down, north, south, east, west}). If the agent's inventory is empty, and the adjacent block contains something, then the agent will take the block, filling its inventory with it. The cube in the world will then be empty. If the agent's inventory is full, then the agent will try to place the block in the world, succeeding if the relevant adjacent cube is empty. The agent can also \textit{do nothing}, or \textit{destroy} the block in its inventory, resulting in an empty inventory. 

This gives the agent 12 possible actions, several of which can reshape the world. To generate behaviour we approximated the empowerment for the states resulting from these 12 actions, and took the action with the highest expected empowerment. In the discrete, noiseless case, empowerment is basically the amount of reachable states, so to fully compute 15-step empowerment we would have to check all $12^{15}$ action sequences, and count how many different results we get. We approximated this by checking 10.000 random sequences, and counting how many different resulting locations (x,y,z-coordinates) the agent would reach. This also provided us with a certain degree of random noise in the action selection. 

\begin{figure}
\centering
\includegraphics[scale=0.3]{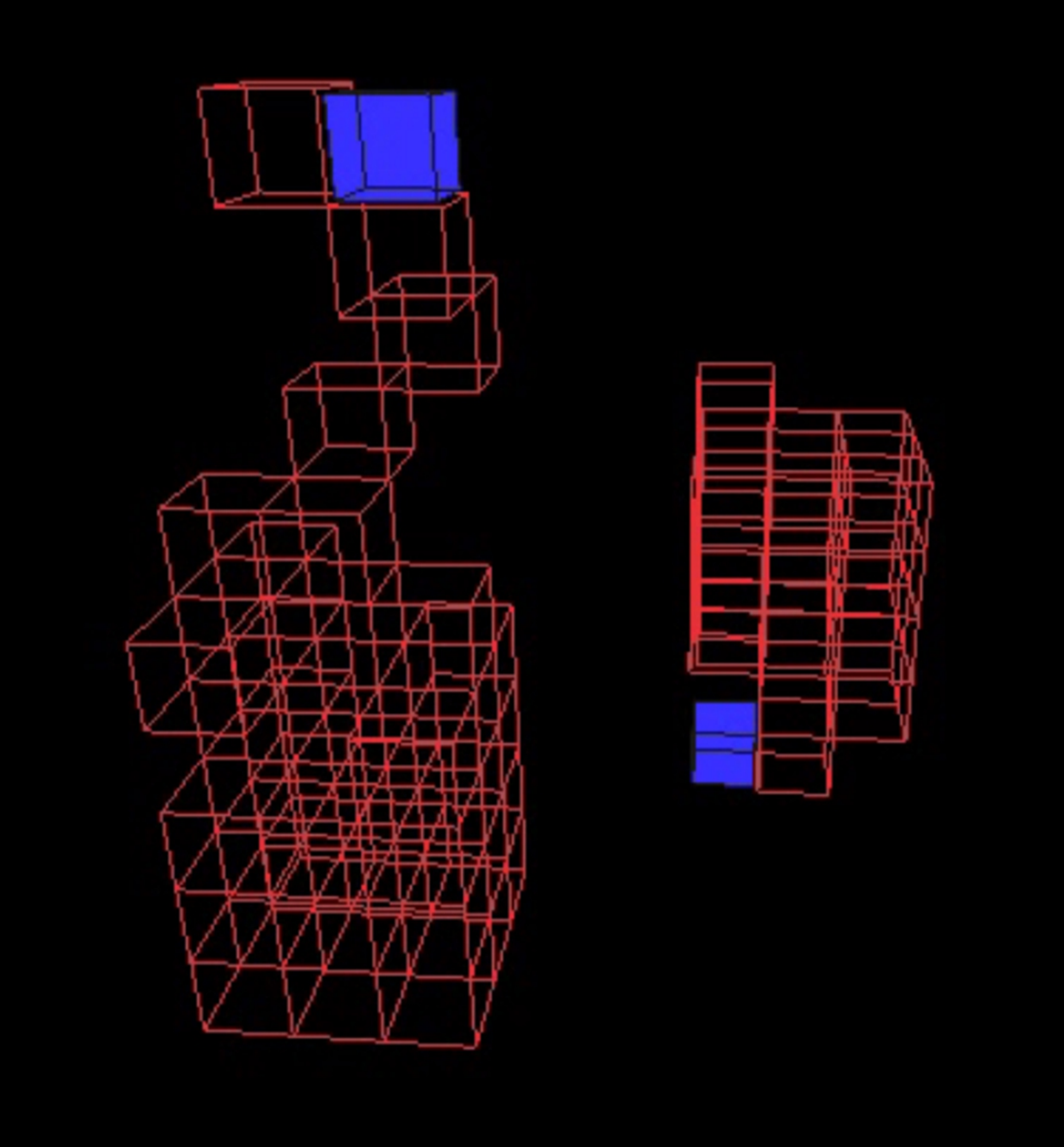}
\caption{A 3d wireframe rendering of two $3 \times 3 \times 10$ worlds, the transparent brown cubes are earth blocks, the blue cubes agents. In the left scenario the agent is controlled by empowerment, and has produced a stair-like structure after ca. 100 time steps, which allows the agent to access the higher parts of the environment. The right scenario features an agent that chooses actions uniformly random for comparison, and we see that the initial configuration of earth blocks in the lower 5 levels is nearly unmodified.}
\label{Fig:1}
\end{figure}

Our agent performs a gradient ascent by choosing the actions that lead to the highest empowerment value in the next step. The observed behaviours depend on the initial configuration of the world. If the world is filled with blocking earth blocks, then the agent will take and destroy blocks to create tunnels through the ground to gain mobility. If the lower half of the world is filled with blocks, then the agent will take in blocks and move them to build stair-like structures (see Fig.~\ref{Fig:1}), to make higher levels of the world accessible to it. The stairs are configured so that the agent-specific movement rules allow it to move up and down in the world; they reflect how the agent interacts with the world, and are artefacts that reflect the agent's motoric affordances. 

In previous simulations \cite{klyubin05:_all_else_being_equal_be_empow} we were able to observe how an agent would change the world state to increase it own empowerment, e.g., an agent would move to the central position in a maze to increase it mobility (and thereby its empowerment). In our current simulation, the agent still changes the worldstate, albeit in a much larger state-space, but the qualitative difference here is that if we remove the agent, and replace it with a similar one in the same world, then the empowerment gain of restructuring the world remains for the second agent. Similarly, in a multi-agent scenario, one agent that is changing the world so it provides more empowerment to itself would at the same time make the world more empowerment-providing for others. Thereby, the intrinsic motivation of empowerment could be used as a local, agent-centric incentive for each member of a group of agents, who would then jointly restructure the world to their common benefit, like a swarm of insects that build a hive together.

\bibliography{GSO2013}
\bibliographystyle{siam}

\end{document}